\documentclass[12pt,preprint]{aastex}

\def\aa#1#2#3#4#5{\bibitem[#1]{#2}#3, {A\&A}, {#4}, #5}

\def\aj#1#2#3#4#5{\bibitem[#1]{#2}#3, {AJ}, {#4}, #5}
\def\apj#1#2#3#4#5{\bibitem[#1]{#2}#3, {Ap. J.}, {#4}, #5}
\def\apjlett#1#2#3#4#5{\bibitem[#1]{#2}#3, {Ap. J. (Letters)}, {#4},
#5}
\def\apjsup#1#2#3#4#5{\bibitem[#1]{#2}#3, ApJS, #4, #5}

\def\icarus#1#2#3#4#5{\bibitem[#1]{#2}#3, {Icarus}, {#4}, #5}

\def\mnras#1#2#3#4#5{\bibitem[#1]{#2}#3, {M.N.R.A.S.}, {#4}, #5}

\def\pasp#1#2#3#4#5{\bibitem[#1]{#2}#3, {PASP}, {#4}, #5}

\begin{document}

\title{Distribution of Circumstellar Disk Masses in the Young Cluster NGC 2024}

\author{J. A.  Eisner and John M. Carpenter}
\affil{California Institute of Technology \\ 
Department of Astronomy MC 105-24 \\
Pasadena, CA 91125}
\email{jae@astro.caltech.edu, jmc@astro.caltech.edu}


%

\begin{abstract}
We determine the distribution of circumstellar disk masses in the
young ($\sim 0.3$ Myr) cluster NGC 2024
by imaging a $2\rlap{.}'5 \times 2\rlap{.}'5$ region
in $\lambda$3mm continuum emission to an RMS noise
level of $\sim 0.75$ mJy beam$^{-1}$ with the Owens Valley Millimeter Array. 
The mosaic encompasses 147 K-band sources as well as the molecular
ridge seen previously in dust continuum emission.
We detect 10 point-like sources in $\lambda$3mm continuum emission
above the level of 5$\sigma$ within the unit gain region of the mosaic.
One of these sources corresponds to the near-IR
source IRS 2, an early B-type star.  Two other sources are tentatively
associated with low-mass near-IR cluster members, and the remaining 7 
sources have no K-band counterparts.  
Assuming the millimeter continuum point sources
represent emission from circumstellar disks and/or envelopes,
then $\sim 6\%$ of the total population (infrared and millimeter sources)
in the NGC 2024 mosaic has a circumstellar mass in excess of
$\sim 0.06$ M$_{\odot}$.
We obtain further constraints on the average circumstellar disk mass
by considering the mean millimeter continuum flux observed toward a sample of 
140 K-band sources that likely have stellar masses $\lesssim 1-2$ M$_{\odot}$. 
While none of these sources are detected individually above the 
3$\sigma$ limit of $\sim 0.035$ M$_{\odot}$, the ensemble of sources
are detected in the mean at the $5\sigma$ level with a mean disk
mass of $\sim 0.005$ M$_{\odot}$.
Compared to the older ($\sim 2$ Myr) cluster IC 348, NGC 2024
contains a higher frequency of massive disks/envelopes and has a 
higher mean disk mass by a factor of $2.5 \pm 1.3$ among K-band
sources, suggesting that the mean circumstellar mass is decreasing with cluster
age.  We also compare the results for the NGC 2024 and IC 348 clusters
to those for the lower-density Taurus star forming region. 
Finally, we compare our detection limits with the minimum
mass estimate for the proto-solar nebula, and discuss possible
implications for planet formation.
\end{abstract}

\section{Introduction}
In recent years, high resolution millimeter, infrared, and optical images
have provided direct evidence for the existence of 
circumstellar disks on scales of $\sim 100$--1000 AU
around young stars \citep[e.g.,][]{KS95,DUTREY+96,
PADGETT+99,OW96}.  These disks are the likely birth-sites for planets,
and determination of the ubiquity, masses, and
evolutionary timescales of circumstellar disks will place
constraints on the timescales and mechanisms of planet formation.

While direct imaging has provided concrete evidence for a
limited number of circumstellar 
disks, high resolution observations are difficult to obtain for large
ensembles of objects. Therefore, indirect tracers of circumstellar disks
are commonly utilized to infer disk evolution. The most common 
tracer is the presence of near-infrared 
(near-IR) emission in excess of that expected from
the stellar photosphere. Indeed, the ubiquity
of near-IR excesses around young stars in Taurus \citep[see, e.g.,][]{STROM+89}
provided early indications that disks are common around young solar-mass 
stars.  More recently, JHKL photometric
surveys of rich clusters spanning a range of ages detected near-IR
excesses for at least 50\% 
of solar mass stars at an age of $\sim 1$ Myr, but for fewer than
10\% of stars with ages from 3--10 Myrs \citep[e.g.,][]{HLL01b}.

The main limitation of near-IR surveys is that the excess emission
traces only the hot ($> 1000$ K), inner ($\sim 0.1$ AU) disk regions around
solar-mass stars.  Moreover, the emission is usually optically thick.
Therefore, near-IR surveys provide no direct constraints on the
reservoir of cold dust located at larger radii where planets are 
expected to form. To study this cooler material, observations at
longer wavelengths, in particular
millimeter (mm) and sub-millimeter (sub-mm) wavelengths, are necessary.
Since the mm and sub-mm emission is optically thin, the 
observed flux provides a direct measure of mass.

Several investigators have carried out 
comprehensive single-dish mm and sub-mm continuum surveys toward
regions of star formation comprising loose aggregates of stars:
Taurus \citep{BECKWITH+90,OB95,MA01}, $\rho$ Ophiuchi 
\citep{AM94,NUERNBERGER+98,
MAN98}, Lupus \citep{NCZ97}, Chamaeleon I \citep{HENNING+93}, Serpens
\citep{TS98}, and MBM 12 \citep{ITOH+03,HOGERHEIJDE+02}.  
About $20$\%--30\% of stars aged
$\sim 1$ Myr appear to possess circumstellar disks with masses 
greater than $\sim 0.01$ M$_{\odot}$, comparable to the minimum mass of the 
pre-solar nebula \citep{WEID+77,HAYASHI81}, and  
the median disk mass is $\lesssim 0.004$ M$_{\odot}$.

Expanding millimeter continuum surveys to include rich
clusters allows the determination of 
accurate statistics on the frequency and evolution 
of disk masses as a function of both stellar mass and age. Also, since most 
stars in the Galaxy form in rich clusters \citep{LADA+91,LADA+93,CARPENTER00}, 
understanding disk formation 
and evolution in cluster environments is a vital component in our general 
understanding of how stars and planets form.
The main challenge to observing rich clusters at mm-wavelengths 
is that very high angular resolution is required to resolve individual sources
and to distinguish compact disk emission from the more extended emission
of the molecular cloud.  Single-aperture mm-wavelength telescopes
lack sufficient angular resolution, and to date, only two rich
clusters have been observed with mm-wavelength interferometers:
the Orion Nebula cluster \citep{MLL95,BALLY+98}, and IC 348 
\citep{CARPENTER02}. These observations have detected 
{\it no} massive disks with 3$\sigma$ upper limits ranging from
$\sim 0.025$--0.17 M$_{\odot}$.

Here, we present a mm-wavelength interferometric survey of
NGC 2024, a young, deeply embedded
stellar cluster in the Orion molecular cloud.  
In addition to improving the statistics on
circumstellar disks in clusters, these observations enable a comparison
between relatively 
young (NGC 2024; $\sim 0.3$ Myr) and old (Orion Nebula cluster and IC 348;
$\sim 1$-2 Myr) clusters, which places constraints on the timescales for disk 
evolution.  In the next section, we describe the NGC 2024 region and discuss
the stellar and protostellar populations.  The observations and results are
presented in \S \ref{sec:obs} and \S \ref{sec:cont}, 
and we derive constraints on the
circumstellar disk masses in \S \ref{sec:disks}.  Finally, we
compare the results for NGC 2024 to those
for the IC 348, Orion Nebula cluster,
and Taurus regions, and discuss the implications for
disk evolution in rich clusters.

\section{The NGC 2024 Cluster\label{sec:ngc2024}}
NGC 2024 is a young HII region embedded in the L1630 (Orion B)
molecular cloud.  
Distance estimates to the Orion region range from $\sim 360-480$ pc
\citep[][and references therein]{A-T82,BGZ94},
and the distance to the stellar group containing NGC 2024 has
been estimated to be 415 pc based on  ubvy photometric and
Balmer line measurements of 11 B-stars \citep{A-T82}.
We adopt this distance of 415 pc.  

Grasdalen (1974) originally identified 
the brightest infrared source in the region, IRS 2, which has
a luminosity of $\sim 10^6$ L$_{\odot}$ and is a suspected early B-type star.
IRS 2 was subsequently resolved into two sources, IRS 2 and IRS 2b
\citep{JPL84}. Recently, Bik et al. (2003) estimated 
a spectral type of $\sim$O8
for IRS 2b, and suggested that it is the dominant source of ionizing
flux for the HII region.

NGC 2024 also contains a cluster
of lower mass stars, originally identified by Barnes (1989).
Lada et al. (1991) identified 309 sources with $m_K < 14$ 
within the NGC 2024 cluster, and Lada (1999) computed a
stellar density of $\rho \approx 400$ stars pc$^{-3}$.
Meyer (1996) obtained J-,H-, and K-band photometry for 233 cluster members.
L-band photometry (and new
JHK measurements) for 142 of these stars by Haisch et al. (2000)
indicates near-IR excess emission 
for $\ge 86 \pm 8$\% of sources.  Thus, a large fraction of the 
NGC 2024 cluster members 
have at least a small amount ($\sim 10^{-6}$ M$_{\odot}$)
of hot ($>1000$ K) associated circumstellar material, presumably distributed
in inner circumstellar disks.

The age of the NGC 2024 cluster has been estimated at $\sim 0.3$--$0.5$ Myr 
{\citep{MEYER96,ALI96}, using
spectroscopic and photometric data to place cluster members in an HR
diagram and then inferring the age from 
pre-main sequence evolutionary tracks by D'Antona \& Mazzitelli (1994).
For comparison, the same technique 
yields ages of $\sim 1$ and 2 Myr
for the Orion Nebula cluster and IC 348, respectively
\citep{HILLENBRAND97,LUHMAN+98,LUHMAN99}.
While the absolute ages of the three clusters are uncertain
due to limitations of the pre-main sequence evolutionary tracks, the relative
ages are more secure.  To further clarify the relative ages of the three
clusters, we re-calculate the ages in a consistent way 
using published data compiled
by Hillenbrand, Meyer \& Carpenter (2003).  We place cluster
members with masses between 0.1 and 1 M$_{\odot}$ on an HR diagram,
and infer the ages from the pre-main
sequence tracks of D'Antona \& Mazzitelli (1997).
We compute logarithmic ages (in years)
of $4.2 \pm 0.9$, $5.5 \pm 0.8$, and $6.2 \pm 0.7$ for
NGC 2024, the Orion Nebula cluster, and IC 348, respectively. 
The relative youth of NGC 2024 is also supported
by the fact that the cluster remains deeply embedded within the molecular 
cloud, in contrast to both the Orion Nebula cluster 
and IC 348 where the extinction to the cluster members is substantially less
\citep{HILLENBRAND97,LUHMAN+98,LUHMAN99}.
In the remainder of the discussion, we adopt an age of 0.3 Myr for
NGC 2024.

Meyer (1996) estimated the masses for NGC 2024 cluster members using
near-IR spectroscopy for 19 sources, and a photometric method based on
de-reddening sources to the expected locus of T Tauri stars in color-color
diagrams for an additional 72 sources.  The resulting mass
distribution is statistically consistent with a Miller-Scalo IMF.
Figure \ref{fig:hk} shows a color-magnitude diagram for NGC 2024
(based on data from Meyer 1996).  Although near-IR excess emission,
age uncertainties, and binarity complicate the interpretation of
this diagram, the observed colors and magnitudes are consistent with most
stars  in NGC 2024 having masses of $\sim 0.1$--1 M$_{\odot}$.
While some sources ($\sim 10\%$) may have sub-stellar
masses, spectroscopic observations are required to confirm this
speculation.

We examined the K-band UKIRT image from Meyer (1996), and found 147
point-like sources within the region we observed at OVRO 
(see \S \ref{sec:obs}).
We astrometrically calibrated the UKIRT image using the 2MASS Point
Source Catalog, and the residuals of the astrometric fit of the UKIRT image
are $0\rlap{.}''1$. The astrometric uncertainty of the 2MASS coordinates is 
$\lesssim 0\rlap{.}''2$, and thus the positions of the infrared sources
are known to within a fraction of the OVRO beam size ($\sim 2''$; see
\S \ref{sec:obs}).

In addition to the young stellar population traced by near-IR emission,
NGC 2024 also contains a population of deeply embedded objects, 
invisible at near-IR wavelengths, but traced by
mm and sub-mm wavelength emission.  Several strong,
compact $\lambda3$mm continuum sources, labeled
FIR 1-7 \citep{MEZGER+88,MEZGER+92}, are detected in the
ridge of mm emission that coincides with the prominent dust lane
in NGC 2024.  While these compact sources were
originally interpreted as dense ($n_{\rm H} \sim 10^{14}-10^{15}$ m$^{-3}$), 
cold ($T \sim 15-20$ K) protostellar condensations
\citep{MEZGER+88, MEZGER+92}, subsequent observations yielded evidence that
at least some of the sources are actually embedded protostars or
young stellar objects \citep[e.g.,][]{MOORE+89,RICHER+89,CC96}.

\section{Observations \label{sec:obs}}
We mosaicked a $2\rlap{.}'5 \times 2\rlap{.}'5$ region toward the
NGC 2024 cluster in the 
$\lambda$3mm continuum with the OVRO millimeter array
between January, 2002 and April, 2003.  
Continuum data were recorded simultaneously in four 1-GHz bandwidth channels
centered at 99.46, 97.96, 102.46, and 103.96 GHz.  
As shown in Figure \ref{fig:pointings},
the mosaic consists of sixteen pointing centers.  Five separate
array configurations provided
baselines between 15 and 483 meters, and  
the uv coverage of the observations is shown in Figure \ref{fig:uv}a.

We calibrated the amplitudes and phases of the data with the
blazar J0530+135: 
$(\alpha,\delta)_{\rm J2000} =(5^{\rm h}30^{\rm m}56\rlap{.}^{\rm s}4, 
+13^{\circ}31'55\rlap{.}''2)$.
We estimated the flux for J0530+135
using Neptune and Uranus as primary flux
calibrators, and 3C84 and 3C273 as secondary calibrators.
Since we obtained data over two separate observing seasons, and since
J0530+135 may be variable, we estimated the flux for each season.
For observations up to May, 2002, we determined a mean
flux of 2.23 Jy, with an RMS
dispersion of 0.16 Jy, 
as computed from observations on 21 separate nights in which both 
J0530+135 and flux calibrators were observed.  For the observations that 
began in September 2002, we computed a mean flux of 1.97 Jy, with an
RMS dispersion of 0.11 Jy, based on observations on 5 nights in which
both J0530+135 and flux calibrators were observed.  
All data reduction and calibration were performed using the OVRO software
package MMA \citep{SCOVILLE+93}.

We mosaicked the sixteen individual pointings into a single image, then
deconvolved and CLEANed the mosaic using
the MIRIAD package \citep{STW95}.  
We averaged the data using robust weighting (with
a robustness parameter of 0.5) in order to obtain a good balance between
sensitivity and angular resolution.  
Based on observations of circumstellar disks in Taurus, which find disk
diameters $\la 300$ AU \citep{DUTREY+96}, we expect that disks
in NGC 2024 (at a distance of 415 pc) will have angular scales of
$\la 0\rlap{.}''7$.  In order to directly observe this compact disk emission,
which is 
effectively point-like at the angular resolution of these observations,
we resolved out the extended emission in the OVRO mosaic by
using only longer-baseline data.
Experimenting with different uv spacing cutoffs, we minimized
the RMS noise of the map by eliminating all
data with uv spacings $< 20$ k$\lambda$.  

Because NGC 2024 is at low
declination, the uv coverage is often sparse in the north-south direction,
and thus beam artifacts in the north-south
direction may contain up to 50\% as much flux as the main beam.
Figures \ref{fig:uv}b and c show the synthesized beams corresponding to the 
full uv coverage of the mosaic, and only uv spacings $> 20$ k$\lambda$,
respectively.  The FWHM of the core of the beam using all our OVRO data is 
$3\rlap{.}''84 \times 3\rlap{.}''22$ at a position angle of $-28.5^{\circ}$,
and the FWHM of the beam core for only outer spacings is
$2\rlap{.}''53 \times 2\rlap{.}''13$ at a position angle of $-44.2^{\circ}$

\section{$\lambda$3mm Continuum Emission \label{sec:cont}}
The mosaic produced from all of our robust-weighted NGC 2024 data
is shown in Figure \ref{fig:map}.  
The unit gain region of the mosaic encompasses a 
$2\rlap{.}'5 \times 2\rlap{.}'5$ area, 
as indicated by the solid contour.
We detect the previously known
mm sources FIR 2-6 \citep{MEZGER+88} and IRS 2 \citep{WMD95}, as well as
extended emission from the molecular ridge
\citep[e.g.,][]{CC96}.
The horizontal emission regions labeled ``NCP'' and ``SCP''
in Figure \ref{fig:map} correspond spatially to the free-free emission
regions observed in VLA centimeter continuum maps
\citep{CRUTCHER+86,GJW92}.   

We calculate the RMS of the image in 
$0\rlap{.}'5 \times 0\rlap{.}'5$ sub-regions 
after removing the strong point source emission using CLEAN.
The ``intrinsic'' RMS noise in the mosaic is $\sim 0.6$ mJy, calculated for
a $0\rlap{.}'5 \times 0\rlap{.}'5$ region in the northwest corner of the map
that is free of any strong emission.  Some regions of the image with more
extended emission and sidelobe artifacts 
(which we could not CLEAN adequately) have an RMS 
as high as $\sim 3$ mJy, and the mean RMS across the mosaic is $1.3$ mJy.

Figure \ref{fig:map-outer} shows the OVRO mosaic produced from data 
with uv spacings $>20$ k$\lambda$.  
The extended emission seen in Figure \ref{fig:map} is mostly resolved out,
and the overall sensitivity is significantly improved.  
Specifically, the RMS sensitivity within a {$0\rlap{.}'5 \times
0\rlap{.}'5$} region in the northwest corner of the
mosaic shown in Figure \ref{fig:map-outer} is $\sim 0.5$ mJy
(compared to $\sim 0.6$ mJy for Figure \ref{fig:map}).
The local RMS in $0\rlap{.}'5 \times 0\rlap{.}'5$ sub-regions
varies from $0.5-1$ mJy across the image, and the mean RMS for the mosaic is
$0.75$ mJy.
Since Figure \ref{fig:map-outer} represents such a dramatic improvement
over Figure \ref{fig:map} in terms of RMS noise and distinctness of
the point sources, we will use this outer spacings map in the
remainder of the analysis.

In Figure \ref{fig:map-outer}, 
we detect emission above the 5$\sigma$ level 
(where sigma is determined locally, as described in the preceding paragraphs)
from the compact sources FIR 2-6 and IRS 2, as well as
several faint new sources.
We choose this detection threshold since
$<0.01$ out of the $\sim 6,000$ independent
pixels within the unit gain contour 
are expected to have noise spikes above the 5$\sigma$ level (assuming
Gaussian noise).
Positions, measured fluxes, sizes, and estimated masses (see
\S \ref{sec:disks} below) for all of the compact sources are listed in
Table \ref{tab:mm}.
Since the extended emission from the NCP and SCP structures is not 
completely resolved out of Figure \ref{fig:map-outer}, no sources observed
towards the NCP and SCP are included.
Since some of the weaker 
sources listed in Table \ref{tab:mm} may represent peaks of
the underlying extended emission or beam artifacts due to imperfect cleaning, 
rather than true point sources, they should be treated with some caution.

For each source in Table \ref{tab:mm}, 
we fit 2-D elliptical Gaussians to the emission to determine fluxes,
positions, and sizes.  For three weak sources, the Gaussian
fits did not converge, and for these we determine the peak flux,
and quote the position of the pixel in which the peak flux occurs.
From the fitted sizes listed in Table \ref{tab:mm}, FIR 2, FIR 3, 
and source \#7
appear to be somewhat resolved.  Since we expect typical disks
in NGC 2024 to be unresolved in the OVRO image
(\S \ref{sec:obs}), these resolved sources
likely represent envelopes, large disks, or other extended structures.

We now restrict our attention to the mm-wavelength fluxes observed towards 
the known K-band cluster members within the unit gain contour.  
For these 147 pre-determined pixel positions, $\sim 0.2$ sources are 
expected to show emission above the 3$\sigma$ level from Gaussian noise,
and we therefore use a 3$\sigma$ detection level of
$2.25$ mJy for individual K-band sources.  We note that 
for the entire mosaic, $\sim 10$
pixels within the unit gain contour are expected to show noise spikes above 
the 3$\sigma$ level, which is why we used a 5$\sigma$ detection limit
above. 

Figure \ref{fig:hist}a shows the distribution of mm-wavelength
fluxes observed towards K-band sources in NGC 2024.  Several K-band
sources have corresponding $\lambda$3mm fluxes above the 3$\sigma$
level.  One of these sources
is IRS 2, an early-type B star previously detected by several investigators 
\citep[e.g.,][]{GRASDALEN74,WMD95}.  
Two other detected sources, FIR 2 and 4, correspond roughly with the
positions of near-IR cluster members.  
However,  the peak of the mm emission for 
FIR 2 is $\sim 1\rlap{.}''6 \pm 0\rlap{.}''3$ away
from the position of the infrared cluster member, and FIR 4 is
$\sim 0\rlap{.}''7 \pm 0\rlap{.}''3$ away from the nearby 
infrared cluster member.  The other source in Figure \ref{fig:hist}a
above the 3$\sigma$ level is part of the NCP extended structure,
and the extreme negative source with $F \sim -4$ mJy is actually tracing
a sidelobe artifact.  We exclude these sources, as well as the
high-mass stars IRS 1 and IRS 2b, in order to examine the distribution of
$\lambda$3mm fluxes observed towards the remaining 140
``typical'' low-mass cluster members.

The observed distribution of $\lambda$3mm fluxes 
shown in Figure \ref{fig:hist}a
is different from that expected for pure
Gaussian noise with a mean of zero and an RMS of 0.75 mJy 
in that the fluxes are skewed 
towards positive values.  Weighting the observed fluxes by the locally 
measured RMS of the image, we determine an
RMS of the distribution of 0.94 mJy, with a mean of 
0.32 mJy, and a standard deviation of the mean of 0.06 mJy.  
This positive bias is significant since the noise in the OVRO map is
consistent with a Gaussian distribution about a mean of zero (see Figure
\ref{fig:hist}b).  Specifically, the distribution of fluxes
observed towards all pixels within the unit gain contour of the OVRO
mosaic (with the strong point sources listed in Table \ref{tab:mm} 
removed using CLEAN) has a mean of $-7.6 \times 10^{-5}$ mJy  
and a standard deviation of $0.78$ mJy.  Comparison of Figures \ref{fig:hist}a
and \ref{fig:hist}b suggests that the bias in the distribution of fluxes
observed towards K-band sources represents real underlying emission.

The positive bias is also illustrated in Figure \ref{fig:avg}, which shows an
``average'' image of the 3mm flux observed towards K-band sources,
obtained by averaging $10'' \times 10''$ images centered around each object.
The bright point sources listed in Table
\ref{tab:mm} were removed using CLEAN before shifting and co-adding the images.
Figures \ref{fig:hist}a and \ref{fig:avg} both exhibit a positive bias
at the $\sim 5\sigma$ confidence level.  Moreover, the FWHM
of the emission in Figure \ref{fig:avg} is consistent with a point source 
centered at the mean position of the K-band sources, 
including negative features that closely resemble the negative features of 
the OVRO beam (Figure \ref{fig:uv}c).  
Based on Figure \ref{fig:avg}, the possibility that the positive
bias in Figure \ref{fig:hist}a is due to extended cloud emission is unlikely
because the average image is centered and is clearly point-like (whereas
we would expect cloud emission to be extended and randomly distributed).
Thus we conclude that the positive bias observed in Figures 
\ref{fig:hist}a and \ref{fig:avg} probably represents underlying
weak mm-wavelength emission from point sources, and that the mean flux
for the ensemble is $0.32 \pm 0.06$ mJy.

\section{Circumstellar Masses \label{sec:disks}}
The $\lambda$3mm continuum sources listed in
Table \ref{tab:mm} represent thermal dust emission, as opposed
to optically-thin
free-free emission from hot plasma, because the $\nu^{-0.1}$ frequency
dependence of free-free emission would imply higher fluxes in
cm-wavelength VLA images from Gaume et al. (1992) than were actually observed.
These authors imaged NGC 2024 at $\lambda$1.3cm with
an RMS noise level of $\sim 0.6$ mJy, and detected emission only from
IRS 2, NCP, and SCP.  For the weakest object in Table \ref{tab:mm},
which has a $\lambda$3mm flux of 
$4.23$ mJy, the implied $\lambda$1.3cm flux for optically-thin
free-free emission would be 4.90 mJy, easily detectable in the Gaume et
al. (1992) image.  In addition, although IRS 2 is detected in the
$\lambda$1.3cm image with $S_{\nu} = 19$ mJy, the measured $\lambda$3mm
flux of 111.17 mJy is much higher than that predicted for optically-thin
free-free emission, indicating that at least some of the $\lambda$3mm
emission is due to dust.

The mass of circumstellar material (dust + gas, assuming a standard ISM
gas to dust ratio) 
can be estimated from the observed flux in the OVRO $\lambda$3mm
continuum image following Hildebrand (1983):
\begin{equation}
M_{\rm circumstellar} = \frac{S_{\nu} d^2}{\kappa_{\nu} B_{\nu}(T_{\rm dust})}.
\label{eq:mass}
\end{equation}
Here, $\nu$ is the observed frequency,
$S_{\nu}$ is the observed flux, $d$ is the distance to the source,
$\kappa_{\nu} = \kappa_0 (\nu / \nu_0)^{\beta}$ is the mass opacity,
$T_{\rm dust}$ is the dust temperature, and $B_{\nu}$ is the Planck function. 
We assume $d=415$ pc \citep{A-T82}, $\kappa_0=0.02$ cm$^{2}$ g$^{-1}$ at 1300 
$\mu$m, $\beta=1.0$ \citep{HILDEBRAND83,BECKWITH+90}, and $T_{\rm dust} = 20$ K
\citep[see discussion in][]{CARPENTER02}.  Uncertainties in the assumed 
values of these parameters (notably $\kappa$) imply that the derived masses are
uncertain (in an absolute sense)
by at least a factor of $3$ \citep[e.g.,][]{POLLACK+94}.
For the sources detected in our mosaic,
we derive circumstellar masses ranging from 
0.07 to 1.71 M$_{\odot}$.  The 5$\sigma$ mass
detection limit is $\sim 0.06$ M$_{\odot}$, although this limit varies
by $\sim 20\%$ across the mosaic.

The OVRO observations alone do not have the angular resolution or
kinematic information necessary to determine whether the
circumstellar material is distributed in circumstellar disks or envelopes,
or combinations of the two.  
If any of our $\lambda$3mm continuum 
sources are also detected in the infrared, 
the observed mm-wavelength dust emission probably arises in flattened
distributions since, for spherical distributions of material the
columns of dust implied by the mm-wavelength flux would completely block
out any near-IR emission \citep[see, e.g.,][]{BECKWITH+90}.  
For even the weakest source
in Table \ref{tab:mm}, the implied extinction for a spherical mass
distribution would exceed $300$ magnitudes.
In contrast, de-reddening the sources in Figure 
\ref{fig:hk} to the 0.3 Myr isochrone yields extinction estimates of 
$\la 50$ magnitudes for the infrared sources in NGC 2024.
Therefore, the millimeter emission detected towards any K-band sources
probably originates from flattened spatial distributions, probably disks,
and not from envelopes of gas and dust. 

NGC 2024 contains 147 K-band detected cluster members and 
10 mm continuum sources within the unit gain region of our OVRO mosaic.
As discussed in \S \ref{sec:cont}, one of the mm sources
corresponds to the K-band source IRS 2, an
early B-type star where the dust temperature could 
be substantially hotter than the 20 K used in Equation \ref{eq:mass}.
Since this would lead to a correspondingly 
lower circumstellar dust mass, the visual extinction argument presented above 
does not necessarily apply, and we cannot rule out circumstellar envelope
emission for IRS 2.  Since the main goal of this study is to determine 
the frequency of disk masses around low mass stars, we exclude IRS 2, 
and by the same argument, we also exclude IRS 1 and IRS 2b from our
analysis \citep{GARRISON68,BIK+03}.
We also exclude two K-band sources that correspond spatially
with mm emission from NCP and a negative sidelobe artifact (\S \ref{sec:cont}).

Of the remaining sample of 151 low-mass objects within the unit gain
contour (9 mm and 142 
near-IR), the mm sources FIR 2 and FIR 4 are tentatively associated with 
K-band cluster members (\S \ref{sec:cont}), leaving 149 unique sources.
Since the associations of K-band sources with FIR 2 and 4 are tentative, 
the quoted fraction of near-infrared sources with disks should be considered an
upper limit. 
None of the remaining 140 K-band sources within the OVRO mosaic, 
which most likely have stellar masses $\la 1$-2 M$_{\odot}$
(see Figure \ref{fig:hk}),
have been detected in the $\lambda$3mm continuum at the locally-determined 
3$\sigma$ noise level or greater.
Therefore, the fraction of low-mass K-band cluster members in NGC 2024 with a 
disk mass $\ga 0.035$ M$_{\odot}$ is at most $1.4\%$.
However, millimeter emission has been detected for the {\it ensemble} of
K-band sources with a mean flux of $0.32 \pm 0.06$ mJy 
(\S \ref{sec:cont}; Figures \ref{fig:hist} and \ref{fig:avg}). 
Since the emission is compact, and centered on the K-band sources,
we suggest that the mm-wavelength
emission originates in disks. Using the assumptions in Equation 
\ref{eq:mass}, this mean flux implies an average circumstellar disk 
mass for the ensemble of low-mass near-IR cluster members 
of  $0.005 \pm 0.001$ M$_{\odot}$.

\section{Discussion \label{sec:disc}}
Excluding the massive stars (see \S \ref{sec:disks}), 
the frequency of circumstellar masses (in disks and/or envelopes) greater than
$0.06$ M$_{\odot}$ in NGC 2024 is $\sim 6\%$ (9/151).
Although K-L color excesses for NGC 2024 cluster members suggest that 
$86 \pm 8\%$ of the stars have circumstellar disks \citep{HLL00}, these
two estimates of the disk fraction are not necessarily contradictory,
since the near-IR emission probes trace
material ($\sim 10^{-6}$ M$_{\odot}$) within $\sim 0.1$ AU of the star, while
millimeter emission traces massive ($\ga 0.035$ M$_{\odot}$) outer
circumstellar material.

We compare our results for NGC 2024 with an analogous
$\lambda$3mm continuum
survey of IC 348 \citep{CARPENTER02}. 
Although the IC 348 survey contains shorter uv spacings than the
NGC 2024 mosaic, and samples $\sim 2$ times larger spatial scales,
this does not affect the comparison since no
emission (compact or extended) was detected in IC 348.
NGC 2024 and IC 348 are similar in that 
each cluster contains on the order of $300$ stars
\citep{LADA+91,HERBIG98}, the 
spectral types of the most massive stars are comparable 
($\sim$O8 for NGC 2024, B0 for IC 348), and the stellar mass distributions 
are consistent with a Miller-Scalo IMF \citep{MEYER96,LUHMAN+98}.
The primary difference between the two clusters
is that NGC 2024 is more deeply embedded within the molecular cloud and 
younger than IC 348 (see \S \ref{sec:ngc2024}).  
No $\lambda$3mm continuum sources were detected in IC 348 with a 
3$\sigma$ upper limit of $0.025$ M$_{\odot}$ out of a sample of 95 known
infrared cluster members.  In contrast, 
$\sim 6\%$ of cluster members 
in NGC 2024 are surrounded by more than $0.06$ M$_{\odot}$ of 
material (which could be distributed in
disks, envelopes, or combinations of the two).
Using the Fisher Exact Test, the
probability that these two measurements are drawn
from the same distribution is $1.5\%$.
Moreover, the average disk mass for the ensemble of low-mass stars
in NGC 2024 is $0.005 \pm 0.001$ M$_{\odot}$, compared to 
$0.002 \pm 0.001$ M$_{\odot}$ in IC 348.  Assuming that the differences
between NGC 2024 and IC 348 are due to temporal evolution, these
observations indicate that massive disks/envelopes dissipate on timescale
$\la 2$ Myr, and that the average disk mass
decreases by a factor of $2.5 \pm 1.3$ between $\sim 0.3$ and 
2 Myr \footnote{Another possible explanation
for the higher observed flux in NGC 2024 relative to IC 348 is that one
or more of the assumed quantities in Equation \ref{eq:mass}
(e.g., temperature or opacity) is different in the two 
regions. Regardless of the underlying factors, our measurement 
suggests disk evolution between $\sim 0.3$ and 2 Myr.}.

The NGC 2024 results can also be compared with the disk masses in
the Orion Nebula cluster, which have been computed from both $\lambda$3.5mm
\citep{MLL95} and $\lambda$1.3mm continuum observations \citep{BALLY+98}.
At $\lambda$3.5mm, a 3$\sigma$ upper limit of 5.7 mJy was derived
for the flux observed above the
level of the expected free-free emission for a sample of 33
proplyds in the Orion Nebula cluster.  Using Equation \ref{eq:mass} with
the same assumptions as in \S \ref{sec:disks}, this
translates into a 3$\sigma$ upper limit on disk mass of 0.17 M$_{\odot}$.
An upper limit on the {\it average} disk mass 
for these 33 sources is 0.03 M$_{\odot}$.
The $\lambda$1.3mm results yield a 3$\sigma$ upper
limit of 0.047 M$_{\odot}$ for five proplyd sources.
The lack of massive disks in the Orion Nebula cluster is similar to
the results for NGC 2024 and IC 348.

In contrast to NGC 2024, IC 348, and the Orion Nebula cluster, 
a large fraction of the stars in Taurus show evidence for massive
circumstellar disks.  
We have compiled a sample
of 164 stars in Taurus that have been observed at millimeter wavelengths,
including sources observed by Beckwith et al. (1990), Osterloh \& 
Beckwith (1995), and Motte \& Andr\'{e} (2001).  All detected sources
included in the sample are $\ge 90\%$ concentrated within the $\sim 11''$ 
beams of the various surveys.   We also compiled K-band
magnitudes for the sample, either from the literature or from 2MASS.
Approximately 21\% of the sample sources show evidence for
circumstellar disks with masses $\ga 0.01$ M$_{\odot}$.

In NGC 2024, $\sim 6\%$ of cluster members have circumstellar masses greater
than $\sim 0.06$ M$_{\odot}$.
In Taurus, the percentage of sources with $>0.06$ M$_{\odot}$ 
of circumstellar material is $\sim 3\%$, and the probability that
the two samples are drawn from the same distribution is 75.9\%.  
Based on these results, we cannot 
statistically distinguish between the frequencies of sources
with circumstellar masses $>0.06$ M$_{\odot}$ in NGC 2024 and Taurus.
However, if we compare the observed mm fluxes for only low-mass K-band sources,
where the millimeter emission likely originates in circumstellar disks,
we find a possible difference.
The fraction of near-IR cluster members in NGC 2024 with circumstellar disks
$\ga 0.035$ M$_{\odot}$ is $\le 1.4\%$. In contrast, 
$\sim 5\%$ of objects in Taurus have circumstellar disks 
$>0.035$ M$_{\odot}$.  
Follow-up observations of the Taurus
sources at high angular resolution show that
the dust usually lies in compact flattened distributions, and not in massive
envelopes \citep[e.g.,][]{DUTREY+96,LOONEY+00}.  Moreover, these objects
are quite bright at K-band, and if placed at the
distance of NGC 2024, would be detectable in the UKIRT 
K-band image for visual extinctions $< 50$ mag. 
Thus, despite the fact that NGC 2024 is younger than Taurus,
the fraction of K-band sources with massive disks is lower. 
The probability that the frequencies of massive disks around K-band
sources in Taurus and NGC 2024 are drawn from the same distribution is
$\le 8.5\%$.

In summary, the NGC 2024 and IC 348 clusters combined contain at most
2 K-band sources out of a total of 239 observed, or $\le 0.8\%$, that are 
associated with millimeter continuum emission characteristic of massive 
circumstellar disks ($M > 0.035$ M$_{\odot}$). In comparison, $5\%$ 
of the sources in Taurus contain such disks. These combined results suggest 
that the rich cluster environments may not be conducive to forming large, 
massive circumstellar disks. Scally \& Clarke (2001) showed that massive 
disks in rich clusters 
may be inhibited by photoevaporation or tidal disruption 
due to close encounters with massive stars, with photoevaporation as the
dominant effect.  Although 
the ionizing flux from the massive stars in 
NGC 2024 will be highly attenuated by the extinction in the dense core,
these effects may help to explain the lack of massive disks.

The derived upper limits on circumstellar disk masses in NGC 2024 and
IC 348 can
be compared with the minimum mass needed to form a system like
our own Solar System.  Summing the mass contained in Solar planets,
and assuming a standard ISM gas to dust ratio, the minimum mass of the
proto-solar nebula is $\sim 0.01$ M$_{\odot}$
\citep{WEID+77,HAYASHI81}.  This reflects primarily the matter
needed to form Jupiter.  The true mass of the
proto-solar nebula may have been higher than this minimum estimate,
depending on the efficiency of conversion of proto-solar dust and gas
into planets.
The average disk mass determined for K-band sources in NGC 2024,
$0.005 \pm 0.001$ M$_{\odot}$, is comparable to the
expected mass of the proto-solar nebula. Although 
uncertainties in the assumed values of the parameters
used in Equation \ref{eq:mass}
imply that the derived masses are
uncertain by at least a factor of $3$,
our data indicate that the average star in NGC 2024 
possesses a disk massive enough to form a planet with $M \la M_{\rm Jupiter}$.

In contrast, the 3$\sigma$ upper limits on disk mass 
in both NGC 2024 and IC 348 rule out the
existence of disks massive enough to form planets with several Jupiter masses.
Our results therefore
imply either that the formation of massive
planets (i.e., several Jupiter masses) is relatively rare in NGC 2024
and IC 348, or that the aggregation of dust grains into planetesimals
(and eventually planets) has already occurred within $\sim 1$ Myr, 
depleting the disks of small grains that OVRO would see in emission.
The timescale for giant planet formation via core accretion is thought to be
$\sim 10^6-10^8$ years \citep[e.g.,][]{POLLACK+96}, and thus it is
unlikely that planets of several Jupiter masses have formed by this method
around most of the stars in NGC 2024 (which has an age of 
$\sim 3 \times 10^5$ years).  
A possible implication of this
might be that giant planets do not form in rich clusters.  A recent
survey of the 47 Tuc cluster failed to detect any planets \citep[where
they expected to find $\sim 17$;][]{GILLILAND+00}, providing
some support for this possibility.
An alternative explanation 
is that the core accretion model for giant planet formation
is not operating in NGC 2024. Giant planet formation through
gravitational instabilities in circumstellar disks requires only
$\sim 10^3-10^5$ years \citep[e.g.,][]{BOSS98}, 
and might provide a viable explanation for
quick massive planet formation in NGC 2024.

\section{Conclusions \label{conc}}
We have imaged the central $2\rlap{.}'5 \times 2\rlap{.}'5$ region of
NGC 2024 in $\lambda$3mm continuum emission with the OVRO
millimeter-wavelength interferometer.  The mosaic encompasses
147 K-band detected cluster members and the molecular ridge seen
previously in dust continuum emission.  We detected 10 point sources
within the unit gain region of the OVRO mosaic above the 5$\sigma$ level 
(where $\sigma \sim 0.75$ Jy).
One of the millimeter sources is coincident with 
the early B-type star IRS 2.   Two other millimeter sources, FIR 2 and FIR 4, 
are near to but not exactly coincident with infrared cluster members.
No millimeter emission was detected towards the other 
low-mass infrared cluster members
above the 3$\sigma$ level of 2.25 mJy.
The mean $\lambda$3mm flux toward the ensemble of K-band sources is
$0.32 \pm 0.06$ mJy.

We use the $\lambda$3mm fluxes to estimate the 
circumstellar dust masses assuming
that the millimeter-wavelength emission is optically thin, with a temperature 
of 20 K, and adopting a mass opacity coefficient of $\kappa=0.02$ cm$^2$ 
g$^{-1}$ at 1.3 mm. With these assumptions, the circumstellar mass 
(dust plus gas) ranges
from  0.07 to 1.71 M$_{\odot}$ for the 10 detected sources. 
The 3$\sigma$ upper limit to
the circumstellar mass around the individual K-band sources (excluding
the tentative associations with FIR 2 and FIR 4) is 
$\sim 0.035$ M$_{\odot}$, and the average mass for the ensemble of low-mass 
K-band sources is $0.005 \pm 0.001$ M$_{\odot}$.
These results show that at the age of the NGC 2024 cluster ($\sim 0.3$ Myr; 
Meyer 1996), $\sim 6\%$ of cluster members have massive ($\ga 0.06$
M$_{\odot}$) circumstellar
structures (disks and/or envelopes), and many of the sources may possess 
low-mass circumstellar disks.

We compare our results to a similar millimeter continuum survey of IC 348
\citep{CARPENTER02}.
None of the 95 cluster members observed in IC 348
were detected in the millimeter continuum above the 3$\sigma$
level of 0.025 M$_{\odot}$ in IC 348.
In contrast, we detect more than $\sim 0.06$
M$_{\odot}$ of circumstellar material around $\sim 6\%$ of cluster members
in NGC 2024.  Moreover, the average disk mass around a typical 
low-mass K-band source in NGC 2024 is $2.5 \pm 1.3$ times higher than
in IC 348.  Thus, there may be some evolution of circumstellar disks and/or
envelopes on $\sim 1$ Myr timescales.

The fraction of circumstellar disks more massive than 
$\sim 0.035$ M$_{\odot}$ around near-IR cluster members
is at most $0.8\%$ in NGC 2024 and IC 348 combined,
suggesting that massive disks are either very rare, or
non-existent, in rich cluster environments.  
In contrast, $\sim 5\%$ of
the sources in Taurus have circumstellar disks more massive than
0.035 M$_{\odot}$, even though Taurus is older than, or of similar age to,
the rich clusters.  This may imply different
physical mechanisms for disk formation and evolution in clustered
versus isolated star forming regions.

The average disk mass of $\sim 0.005$ M$_{\odot}$ for
the ensemble of K-band sources is comparable to the 
minimum mass of material necessary to form the proto-solar nebula,
which implies that many objects in NGC 2024
possess disks massive enough to form planets with the approximate mass of
Jupiter.  However, the 3$\sigma$ limit of 0.035 M$_{\odot}$ on 
the masses of circumstellar disks in NGC 2024 suggests that
high-mass planets (several Jupiter masses) may have either 
already formed (thus depleting the disks of small grains), or may
never form in this cluster.

\noindent{\bf Acknowledgments.}  JAE is supported by a Michelson Graduate
Research Fellowship.
JMC acknowledges support from Long Term Space Astrophysics Grant NAG5-8217 and
the Owens Valley Radio Observatory, which is supported by the National Science
Foundation through grant AST-9981546.
This publication makes use of data products from the Two Micron
All Sky Survey, which is a joint project of the University of Massachusetts
and the Infrared Processing and Analysis Center, funded by the National
Aeronautics and Space Administration and the National Science Foundation.
2MASS science data and information services were provided by the Infrared
Science Archive (IRSA) at IPAC.

\clearpage
\begin{deluxetable}{lcllccccl}
\rotate
\tablewidth{0pt}
\tablecaption{Sources detected in $\lambda$3mm continuum with OVRO 
\label{tab:mm}}
\tablehead{\colhead{ID} & \colhead{Source Name} & \colhead{$\alpha$ (J2000)} & 
\colhead{$\delta$ (J2000)} & \colhead{FWHM} & \colhead{PA ($^{\circ}$)}&
\colhead{$S_{\nu}$ (mJy)$^{a}$} & 
\colhead{M$_{\rm circumstellar}$ (M$_{\odot}$)} & \colhead{}}
\startdata
1 & FIR 2 & 05 41 42.59 & -01 54 9.21 & $3\rlap{.}''43  \times 2\rlap{.}''39$ &
 -49.5 & $5.07 \pm 0.76$ & 0.08 & \\
2 & ... & 05 41 42.83 & -01 54 14.48 & & & $6.19 \pm 0.76$ & 0.10 & \\
3 & FIR 3 & 05 41 43.11 & -01 54 26.15 & $3\rlap{.}''63  \times 2\rlap{.}''26$ 
 & -34.5 & $34.22 \pm 0.93$ & 0.53 & \\
4 & FIR 4 & 05 41 44.13 & -01 54 45.58 & $2\rlap{.}''10  \times 2\rlap{.}''80$ 
 & -143.1 & $15.41 \pm 0.86$ & 0.24 & \\ 
5 & ... & 05 41 44.23 & -01 55 2.72 & $2\rlap{.}''42  \times 1\rlap{.}''91$ &
 -0.5 & $4.26 \pm 0.73$ & 0.07 & $\ast$ \\
6 & ... & 05 41 44.67 & -01 55 1.83 & & & $4.82 \pm 0.73$ & 0.07 & $\ast$ \\
7 & ... & 05 41 45.04 & -01 55 4.33 & $3\rlap{.}''23  \times 2\rlap{.}''41$ &
 -20.9 & $4.23 \pm 0.73$ & 0.07 &  \\
8 & ... & 05 41 45.41 & -01 54 26.39 & $2\rlap{.}''54  \times 2\rlap{.}''35$ &
 24.8 & $15.23 \pm 1.10$ & 0.24 & \\
9 & IRS 2 & 05 41 45.80 & -01 54 29.71 & $1\rlap{.}''94  \times 2\rlap{.}''17$
 & 36.5 & $111.17 \pm 1.10$  & 1.71 & \\
10 & ... & 05 41 45.83 & -01 53 49.37 & $2\rlap{.}''25  \times 2\rlap{.}''88$ &
 5.9 & $5.93 \pm 0.88$ & 0.09 & \\
\hline
11 & FIR 5 & 05 41 44.33 & -01 55 40.98 & 
$2\rlap{.}''79  \times 2\rlap{.}''47$ 
& -164.5 & $81.97 \pm 0.87$ & 1.27 & \dag \\
12 & ... & 05 41 44.48 & -01 55 41.33 & & & $16.67 \pm 0.87$ & 0.27 & \dag \\
13 & FIR 6 & 05 41 45.14 & -01 56 4.22 & $2\rlap{.}''54  \times 2\rlap{.}''06$ 
 & -40.1 & $58.55 \pm 1.32$ & 0.90 & \dag \\

\enddata
\tablerefs{\dag--These sources lie outside of the unit gain contour, 
and have been scaled by the inverse gain; 
$\ast$--These sources are probably artifacts due to imperfect
cleaning of extended emission; $a$--Uncertainties are 1$\sigma$, where
$\sigma$ is the locally-determined RMS.
.}
\end{deluxetable}

\begin{figure}
\plotone{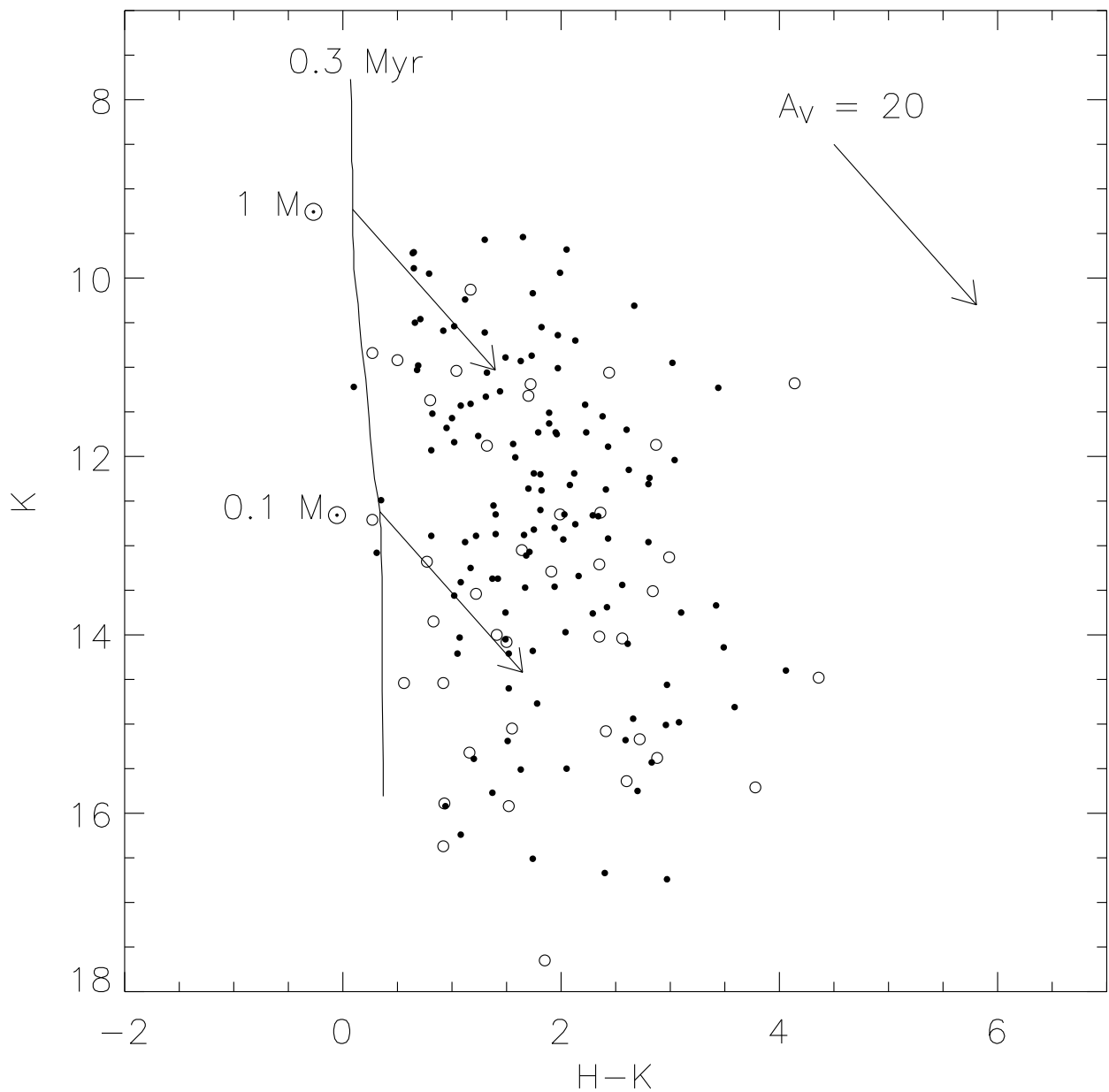}
\caption{Color-magnitude diagram for the stars in the NGC 2024 cluster
(based on data from Meyer 1996).
We also plot the isochrone for stars of age $0.3$ Myr \citep{DM97}, and the
extinction vectors \citep{COHEN+81} for a 1 M$_{\odot}$ star and a 0.1
M$_{\odot}$ star, assuming $A_V = 20$.  Sources that lie within the
unit gain contour of our OVRO mosaic (see Figure \ref{fig:pointings})
are indicated by filled circles, while open circles represent cluster
members that lie outside of the unit gain region.
\label{fig:hk}}
\end{figure}

\begin{figure}
\plotone{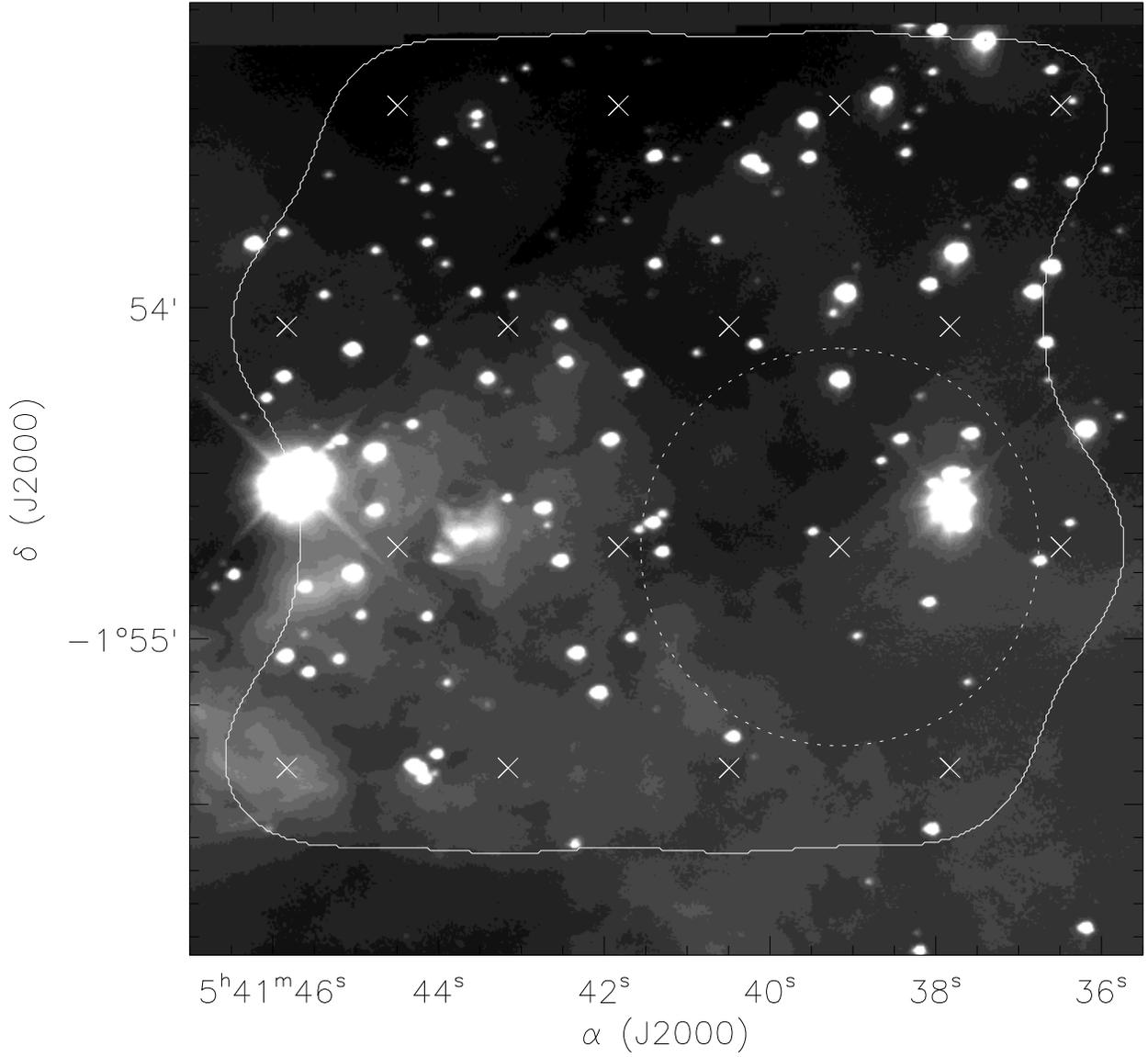}
\caption{Pointing positions for the OVRO mosaic (``X'' symbols), 
plotted over a K-band
image of the NGC 2024 cluster from Meyer (1996).  
The FWHM of the OVRO primary beam at the observed frequency
(100 GHz) is indicated by the 
dotted circle in the lower-right
corner, and the unit gain contour of the map is shown by the solid curve.
\label{fig:pointings}}
\end{figure}

\begin{figure}
\plotone{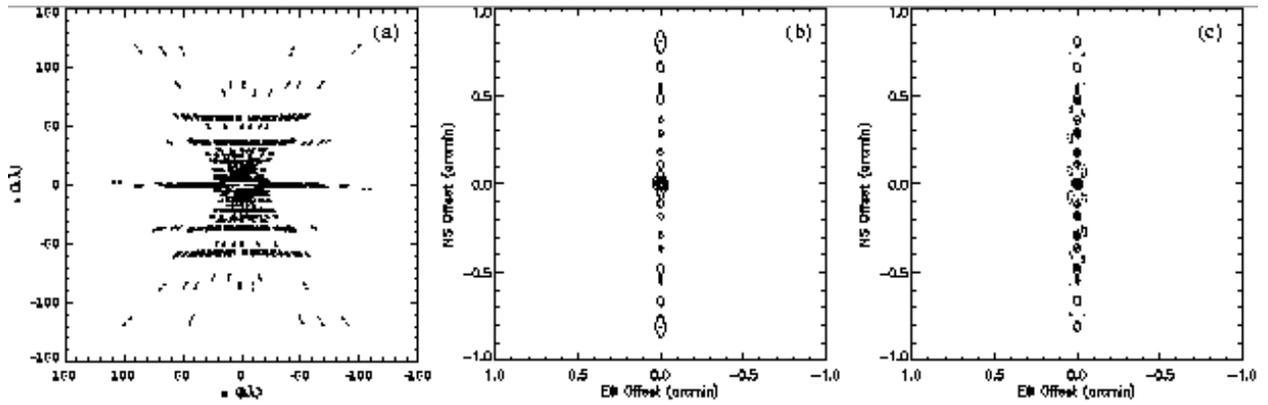}
\caption{(a) uv coverage of the OVRO observations of NGC 2024.
(b) Resultant beam after weighting the complete OVRO data set using a robust
parameter of 0.5. (c) The OVRO beam for only uv spacings $>20$ k$\lambda$.
The contour increments in panels (b) and (c) are 0.2, positive contours are
indicated by solid lines, and negative contours are indicated by dotted
lines.
\label{fig:uv}}
\end{figure}

\begin{figure}
\plotone{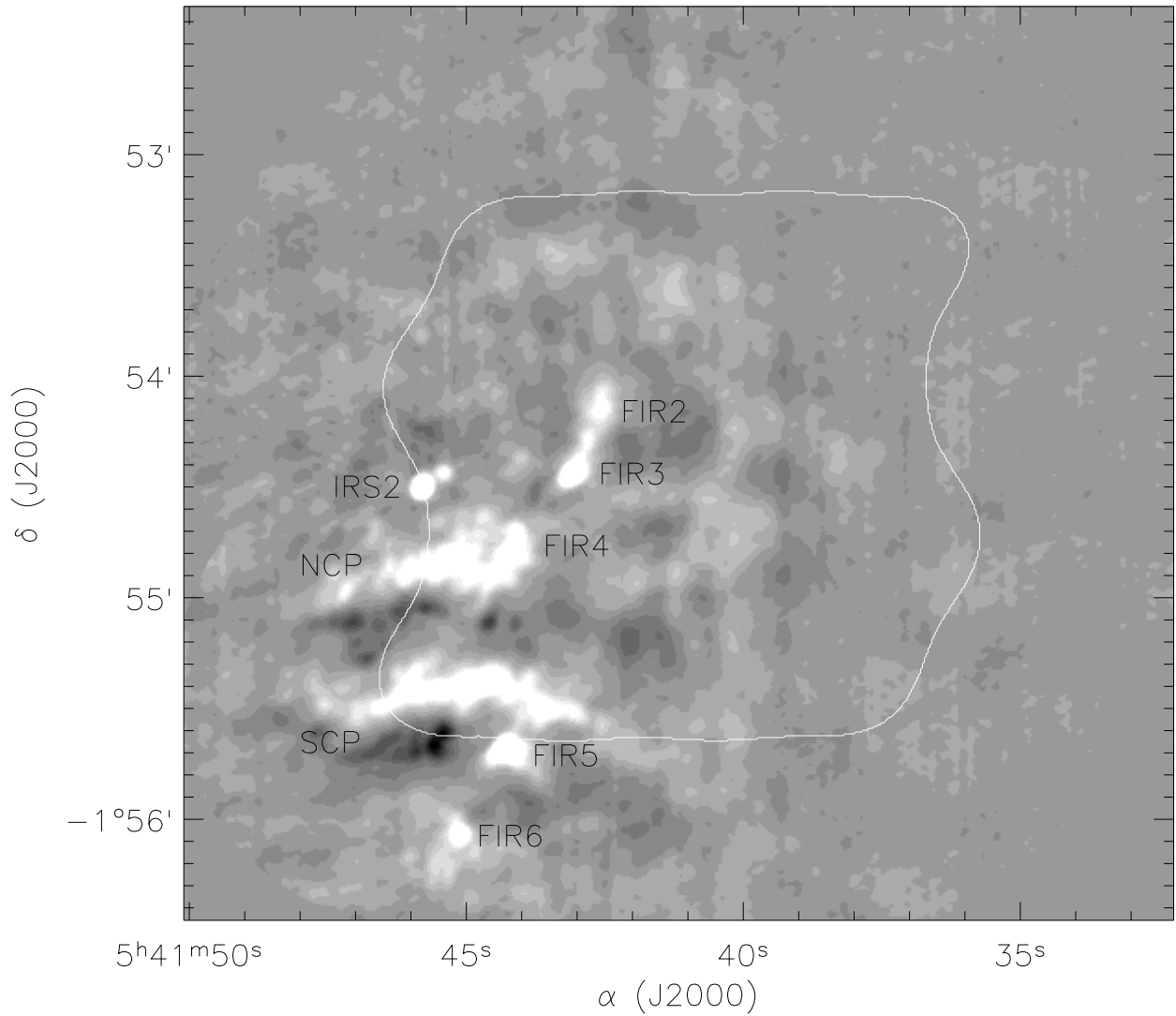}
\caption{The NGC 2024 star forming region, imaged in $\lambda$3mm
continuum with the
Owens Valley Millimeter Array (greyscale).  All of the OVRO data were
used to create this image, and the
angular resolution is $3\rlap{.}''84 \times 3\rlap{.}''22$.
The unit gain region of the mosaic encompasses a 
$2\rlap{.}'5 \times 2\rlap{.}'5$ area, as indicated by the solid contour, and
the average RMS of the residuals within the unit gain contour is 
$\sim 1.3$ mJy. We have labeled the previously detected sub-mm sources
FIR 2-6 \citep{MEZGER+88} and IRS 2 \citep{WMD95}. 
Also, the horizontal emission regions labeled NCP and SCP correspond 
spatially to the free-free emission regions observed in VLA centimeter
continuum maps \citep{CRUTCHER+86,GJW92}.  
While it is difficult to distinguish FIR 4
from NCP in this image, FIR 4 is clearly visible in Figure \ref{fig:map-outer}.
\label{fig:map}}
\end{figure}

\begin{figure}
\plotone{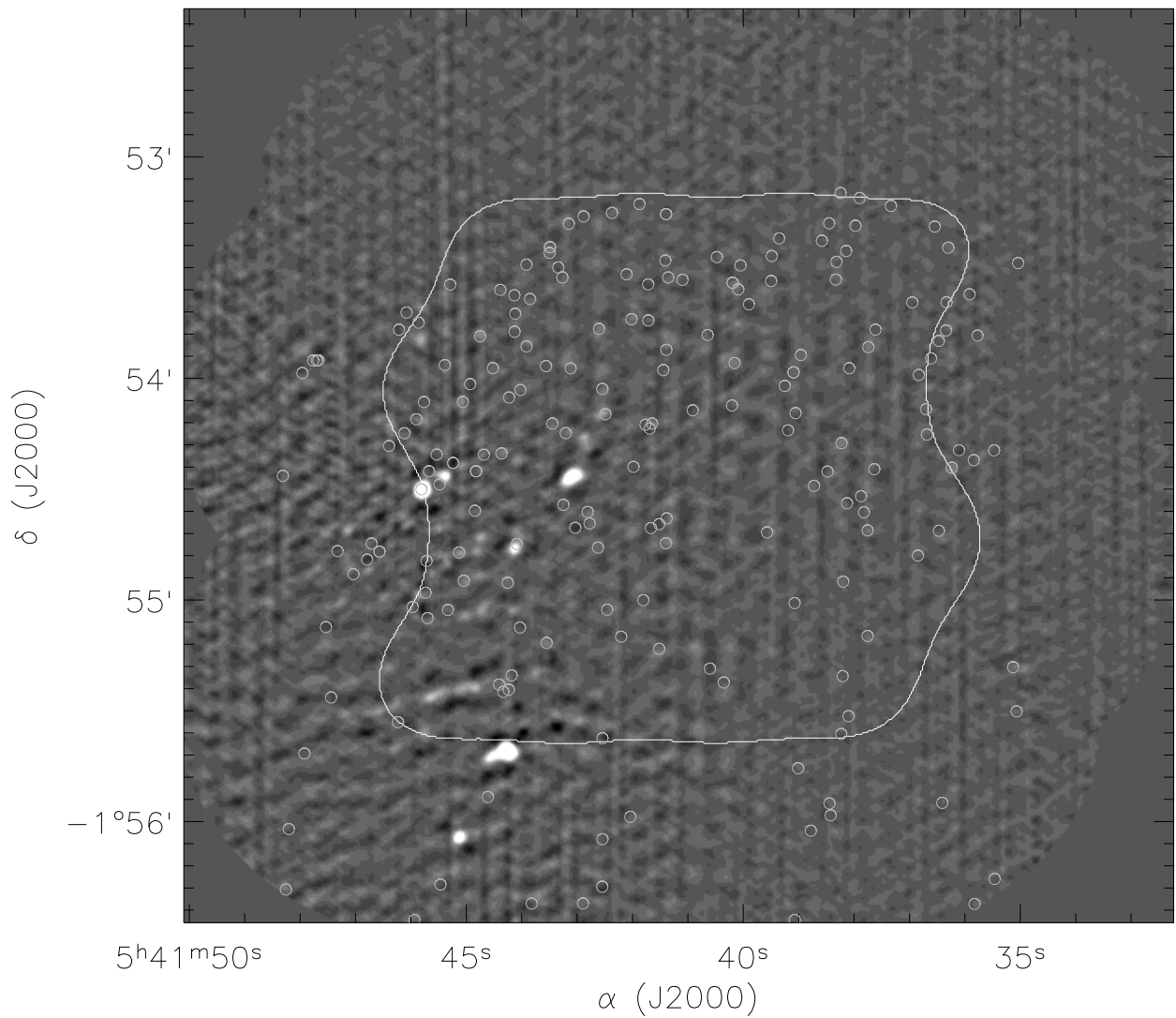}
\caption{The NGC 2024 star forming region, imaged in $\lambda$3mm
continuum with the
Owens Valley Millimeter Array (greyscale).  
This image is constructed from
measurements with a uv radius $>20$ k$\lambda$, and the resultant
angular resolution is $2\rlap{.}''53 \times 2\rlap{.}''13$. 
The unit gain contour is
indicated by a solid line, and
the average RMS of the residuals 
within the unit gain contour is 0.75 mJy. 
The positions of near-IR sources detected at K-band 
\citep{MEYER96} are indicated 
with open circles.
\label{fig:map-outer}}
\end{figure}

\begin{figure}
\plotone{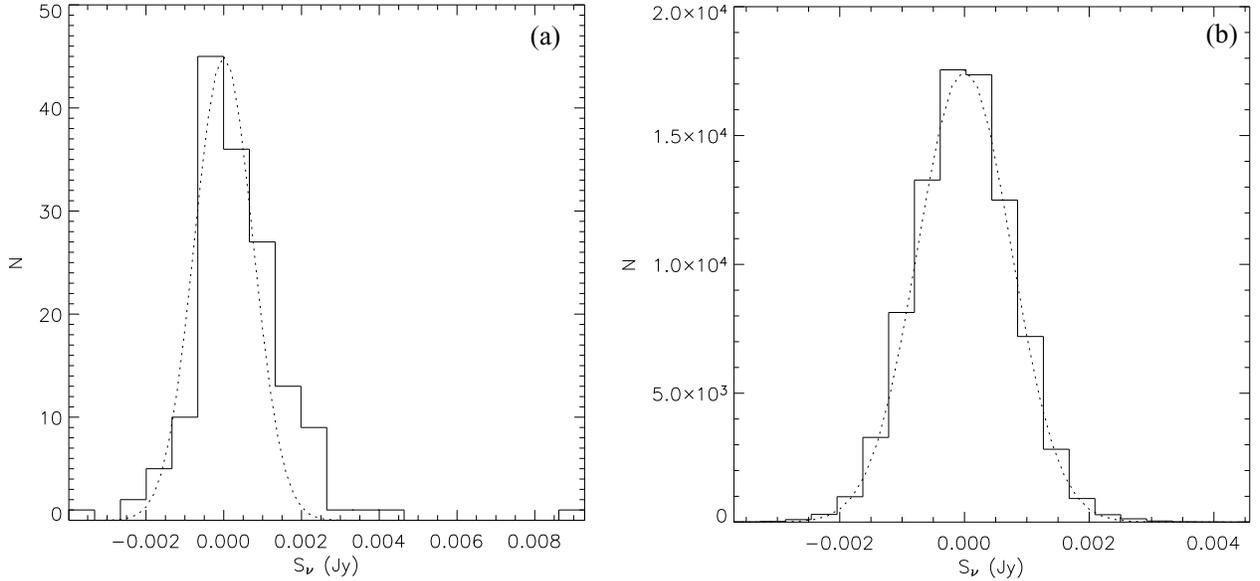}
\caption{(a) Frequency distribution of $\lambda$3mm continuum fluxes
measured towards the low-mass NGC 2024 K-band cluster members 
within the unit
gain contour of the OVRO mosaic (histogram).  
IRS 2 is off the scale of the plot at $\sim 0.1$ Jy.
The three sources with fluxes
$\ga 2.5$ mJy correspond to slight offsets from FIR 2 and 4
(\S \ref{sec:cont}) and to part of the NCP structure.  
The extreme negative source with $F \sim -4$ mJy is actually tracing
a sidelobe artifact. 
(b) Frequency distribution of $\lambda$3mm continuum fluxes for all
pixels within the unit gain contour of the high-angular resolution
OVRO map (Figure \ref{fig:map-outer}), where the bright ($>3.75$ mJy) 
point sources listed in Table \ref{tab:mm} 
have been removed using CLEAN (histogram).
In both panels, the frequency distribution expected for Gaussian noise with
a mean of zero and an RMS of 0.75 mJy is indicated by a dotted line.
\label{fig:hist}}
\end{figure}

\epsscale{0.8}
\begin{figure}
\plotone{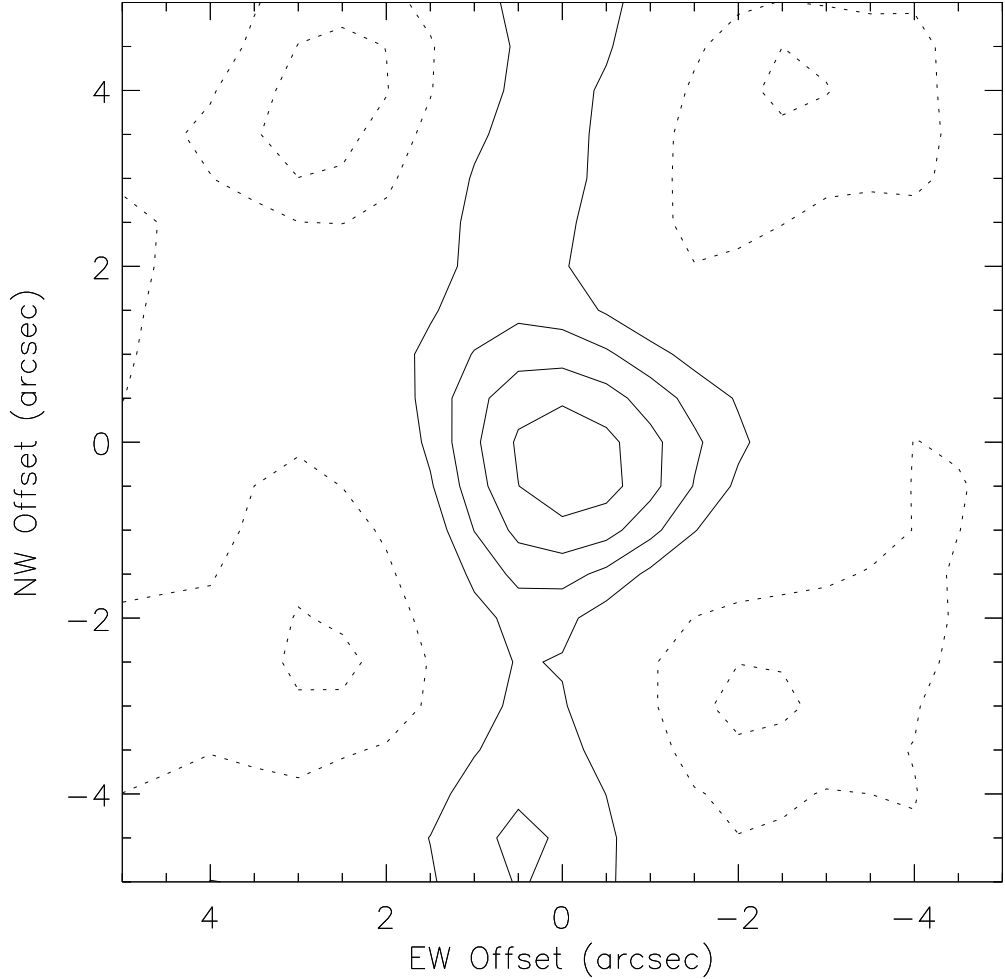}
\caption{OVRO image created by averaging
$10'' \times 10''$ images towards the positions
of known low-mass K-band cluster members in NGC 2024.
The bright sources listed in Table \ref{tab:mm} were removed
using CLEAN before shifting and co-adding the images.
The contour increment is given by the standard deviation of the
mean for the flux distribution (0.06 mJy; \S \ref{sec:cont}). 
Negative contours are
represented by dotted lines and positive contours are represented
by solid lines.  The FWHM of the emission, as well as the
negative features, are consistent with the OVRO beam shown in
Figure \ref{fig:uv}c.  We also note that the emission is centered
at (0,0), corresponding to the mean position of K-band sources.
None of the individual stars used to create this 
image were detected at the $\ge 3\sigma$ level, although the
sources are detected in the mean at the $\sim 5\sigma$ level.
Assuming this compact emission originates from circumstellar disks,
the average disk mass for the ensemble of K-band sources is
$\sim 0.005$ M$_{\odot}$.
\label{fig:avg}}
\end{figure}

\end{document}